\documentstyle[12pt,psfig]{article}
\begin{document}
\title{ZPC 1.0.1:
a parton cascade for ultrarelativistic heavy ion collisions
\footnote{This work was supported by the Director, Office of Energy Research,
Division of Nuclear Physics of the Office of High Energy and Nuclear Physics of
the U.S. Department of Energy under contract No. DE-FG02-93ER40764}
}
\author{Bin Zhang\\
{\em Physics Department, Columbia University,}\\
{\em New York, NY 10027, USA}}
\date{}
\maketitle

\begin{abstract}
\noindent A Monte Carlo program solving Boltzmann equation for
partons via cascade method is presented. At presented, only gluon-gluon elastic
scattering is included. The scattering cross section is regulated by a medium
generated screening mass. Three different geometric modes (3 dimension
expansion, 1-d expansion, and scattering inside a box) are provided for
theoretical study of the parton transport and the applicability of the cascade
method. Space cell division is available to save the number of computer
operations. This improves the speed of the calculation by a large factor and
makes the code best optimized for simulation of parton cascade in
ultrarelativistic heavy ion collisions. 
\end{abstract}

\newpage

\begin{center}
{\Large \bf Program Summary}\\
\end{center}

{\noindent\em Title of program:} ZPC 1.0.1\\

{\noindent\em Program obtainable from:} CPC Program Library, Queen's University
of Belfast, N. Ireland and from the author at bzhang@nt1.phys.columbia.edu\\

{\noindent\em Computer for which the program is designed:} SPARC stations;
{\em Installation:} Physics Department, Columbia University, USA\\

{\noindent\em Computers on which it has been tested:} (1) IBM RS/6000; (2) DEC
Personal Workstation 600AU; 
{\em Installation:} (1) RHIC Computing Facility, Brookhaven National
Laboratory, USA; (2) Theory Program, Nuclear Science Division, Lawrence
Berkeley Laboratory, USA\\

{\noindent\em Operating systems under which the program has been tested:}
SunOS 4.1.3, AIX 1\\

{\noindent\em Programming language used:} FORTRAN 77\\

{\noindent\em Memory required to execute with typical data:} 1,450 words\\ 

{\noindent\em No. of bits in a word:} 32\\

{\noindent\em Number of lines in distributed program:} 6330\\


{\noindent\em Nature of physical problem}

\noindent In ultrarelativistic heavy ion collisions, perturbative QCD predicts
the production of a large number of minijet gluons. The gluon system produced
near central rapidity can be considered as an almost longitudinal boost
invariant and baryon free system. Gluons inside this hot matter will interact
among themselves before hadronization. The effect of the final parton
interaction on the global variables should be studied to match the experimental
observables to the pQCD predictions.\\

{\noindent\em Method of solution}

\noindent The space time evolution of final state partons can be approximated
by a Boltzmann equation. To solve the Boltzmann equation, we employ the cascade
method in which two partons scatter when their closest distance in a specified
frame is less than the interaction length. Gluon scattering cross section is
regulated by a medium generated screening mass. Space cell division is made to
speed up the calculation. Options to do the cascade with
and without space cell division are provided to check the applicability of the
optimization.\\

{\noindent\em Typical running time}

\noindent The running time depends on the density of partons and the scattering
cross section. For 3 dimensional expansion, when the initial density in the
local rest frame is $\approx 50/fm^3$ (at proper time $0.1\;fm/c$, with
transverse radius $5\;fm$), and the parton
parton scattering cross section is $3.5\;mb$, with space cell division, it
takes around 6 minutes to finish 1 event ($\approx 7000$ binary collisions) on
SPARC20, and 30 seconds on alpha Personal Workstation.\\

{\noindent\em Unusual features of the program}

\noindent none

\newpage

\begin{center}
{\Large \bf Long Write-Up}\\
\end{center}

\section{Introduction}

In ultrarelativistic heavy ion collisions at collider energies, hard and
semihard processes dominant. The high energy density produced in these
collisions opens the possibility of studying many interesting
phenomena of strong interactions, e.g., deconfinement phase transition and 
chiral symmetry restoration. The space time evolution of the hot matter
produced in the collisions may make the final observables drastically 
\cite{ruuskanen1} different from those directly predicted from pQCD and Glauber
geometry. So it's very important to study the final state interactions in order
to link the experimental observables and the pQCD predictions. 

The final state parton interactions can be described by the equations of motion
for the quark and gluon Wigner operators. Under semiclassical approximation and
the Abelian dominance approximation, in the weak field limit, the equations of
motion become a set of equations resemble the classical transport
equations\cite{elze1}. This motivates the study of final state interaction by
solving the Boltzmann equation for quarks and gluons:
\begin{eqnarray}
p^\mu\partial_\mu f_a(\vec{x},\vec{p},t)=
& \sum_m\sum_{b_1,b_2,\cdots,b_m}
\int\prod_{i=1}^m\frac{d^3p_{b_i}}{(2\pi)^32E_{b_i}}
f_{b_i}(\vec{x},\vec{p_{b_i}},t)
\nonumber \\
& \sum_n\sum_{c_1,c_2,\cdots,c_n}
\int\prod_{j=1}^n\frac{d^3p_{c_j}}{(2\pi)^32E_{c_j}}|M_{m\rightarrow n}|^2
\nonumber \\
& (2\pi)^4\delta^4\left(\sum_{k=1}^mp_{b_k}-\sum_{l=1}^np_{c_l}\right)
\nonumber \\
& \left[-\sum_{q=1}^m\delta_{ab_q}\delta^3(\vec{p}-\vec{p_{b_q}})
+\sum_{r=1}^n\delta_{ac_r}\delta^3(\vec{p}-\vec{p_{c_r}})\right].
\end{eqnarray}

In the above equation, $f_a(\vec{x},\vec{p},t)$ is the phase space distribution
of parton type $a$, and $M_{m\rightarrow n}$ is the scattering amplitude
defined by: 
\[<c_1,c_2,\cdots,c_n|S|b_1,b_2,\cdots,b_m>
=M_{m\rightarrow n}(2\pi)^4\delta^4\left(\sum_{k=1}^mp_{b_k}-\sum_{l=1}^np_{c_l}\right)\]

To study the final state parton interactions, we have developed ZPC, which
solves the Boltzmann equations for partons by the cascade method.
In the cascade model \cite{cascade1}, two partons scatter
when their closest distance is less than $\sqrt{\frac{\sigma}{\pi}}$ ($\sigma$
is the parton parton scattering cross section). The scattering angle is
determined by the differential parton parton scattering cross section
\cite{combridge1}. Because cascade prescription of solving the semiclassical
transport equation doesn't assume any particular form of the phase space
distribution, it can be used to study systems far from local thermal
equilibrium and it takes into account of finite mean free path effect
automatically.

In the inside outside cascade picture, the partons 
come out as independent partons after the nucleus-nucleus collision. These
partons are taken as the initial condition for the parton cascade.
The parton formation proper time usually depends on the transverse momentum of 
the parton \cite{gyulassy1}. In the simplest picture, the partons are formed
on a hyperbola  of constant proper time \cite{bjorken1} determined by the
average transverse momentum. 

The space time evolution afterwards depends also on the scattering cross
section. Currently, a rough Yukawa type of screening is
used to regulate the parton scattering cross section.
Lattice gauge theory \cite{gao1} and semiclassical transport theory
\cite{biro1} can give us screening mass(es) for parton interactions for
equilibrium system or system close to local thermal equilibrium. Assume that
the formula can be used for the far from equilibrium case as in the
ultrarelativistic heavy ion collisions, we can get an estimate for the medium
generated screen mass. On
other hand, this provides one possible way of measuring the effective
screening in the medium from the experiment.

In ultrarelativistic heavy ion collisions, there are indications that
multiparton processes other than elastic scattering may dominate \cite{xiong1}.
Also because inelastic collisions are important for the chemical equilibrium,
it is important to look at parton production and annihilation processes. 
Besides inelastic collisions, parton statistics and hadronization may all 
affect the prediction in a non-negligible way. These elements will be
incorporated in ZPC in the future.

ZPC is developed in parallel with GCP by Yang Pang \cite{cascade1}. The major
difference is that ZPC 1.0.1
focuses on gluon elastic scatterings, while GCP is being designed to be a more
general, user friendly package. ZPC 1.0.1 has the local interaction list
optimization. ZPC 1.0.1 is written in FORTRAN, while GCP is written in C.

In this paper, we will discuss the physics background for the parton cascade
simulations, some ZPC technical points, then a detailed description of the
program followed by the instructions for users.

\section{Cascade simulations of URHICs}

For Ultrarelativistic Heavy Ion Collisions (URHICs) at RHIC energies and
beyond, the gluon number is much larger than the sea quark and antiquark
number, and the gluon gluon elastic cross section is much
larger than the gluon-quark transition cross section \cite{xiong1}. To first
approximation, we consider only gluon gluon elastic scatterings. The initial
conditions can be read from an external file or some ideal initial conditions
can be generated within the code. 

\subsection{Formation time}

The formation time of partons gives the time beyond which the radiated partons
can be considered as independent partons. There are different models for the
formation time. A simple model motivated by the Uncertainty Principle and used
very often for ideal theoretical calculations is proposed by
Bjorken\cite{bjorken1}. In this model, partons are formed on a hyperbola of
constant proper time. The integrated formation probability can be written as:
\[P(\tau)=\theta(\tau-\tau_0).\]
$\tau_0$ is inversely proportional to $<m_T>$, or can be simply taken as
$1/<m_T>$. 
Another more realistic model proposed by Gyulassy, Wang, and Pl\"umer
\cite{gyulassy1}has an integrated probability:
\[P(\tau)=1/(1+(\tau_f/\tau)^2),\]
where $\tau_f=cosh\;y/m_T$. The above formula interpolated between the
perturbative result $P\propto \tau^2$ for small times and the asymptotic result
which is characterized by a formation time $cosh\;y/m_T$.

In ZPC 1.0.1, the ideal initial conditions are generated with a fixed formation
proper time. Options are available to generate formation time for parton
configurations generated at $t=0$ and propagate the partons to their
formation time.

\subsection{scattering cross section}

The gluon-gluon elastic scattering differential cross section is:
\[\frac{d\sigma}{dt}=\frac{\pi\alpha_S^2}{s^2}\frac{9}{2}
\left(3 - \frac{ut}{s^2} - \frac{us}{t^2} - \frac{st}{u^2}\right).\]
At fixed $s$, the leading divergent term is:
\[\frac{d\sigma}{dt}=\frac{9\pi\alpha_S^2}{2s^2}
\left(\frac{1}{t^2}+\frac{1}{u^2}\right).\]
Since two outgoing particles are identical particles, the scattering angle goes
from $0$ to $\frac{\pi}{2}$. If we let the scattering angle go
from $0$ to $\pi$, then the $t$, $u$ singular terms are redundant. Hence,
\[\frac{d\sigma}{dt}=\frac{9\pi\alpha_S^2}{2s^2}\frac{1}{t^2}.\]
The singularity is regularized by a medium generated screening mass. We end up
with a scattering differential cross section of the form:
\[\frac{d\sigma}{dt}=\frac{9\pi\alpha_S^2}{2s^2}\frac{1}{(t-\mu)^2}.\]
This is just the Yukawa type of scattering cross section. Currently, we set the
total cross section to be $s$ independent:
\[\sigma=\frac{9\pi\alpha_S^2}{2\mu^2},\]
and the differential cross section changes correspondingly to:
\[\frac{d\sigma}{dt}=\frac{9\pi\alpha_S^2}{2s^2}\left(1+\frac{\mu^2}{s}\right)
\frac{1}{(t-\mu)^2}.\]
This is only a simplified model for the gluon-gluon scatterings. Following the
instructions for users, it's easy to make the cross section more realistic.

\subsection{Screening mass}

When the gluon system is not far from equilibrium, perturbative calculation
relates the electric screening mass $m$ with the phase space density
$f(\vec{k})$ through:
\[m^2 = - \frac{3\alpha_S}{\pi^2}lim_{|\vec{q}|\rightarrow 0}
\int d^3k \frac{|\vec{k}|}{\vec{q}\cdot\vec{k}}\vec{q}\cdot
\vec{\nabla}_{\vec{k}}f(\vec{k}).\]

Assuming the formula can be used for nonequilibrium as well, the phase space
density $f(\vec{k})$ is related to the gluon invariant momentum distribution
$g(y,k_T)=d^2N_G/dydk^2_T$ through:
\[f(\vec{k})=\frac{2(2\pi)^2}{g_GV}\frac{1}{|\vec{k}|}g(y,k_T)\]
At collider energies, the transverse screening mass is almost the same as the
longitudinal screening mass. In the ideal case that the rapidity distribution
is an inverted square well and the transverse momentum distribution is an
exponential, the screening mass can be estimated through:
\[m_T^2\approx \frac{3\pi\alpha_S}{R_A^2}\frac{N_G}{2Y}.\]
With $\alpha\sim 0.4$, $\frac{N_G}{2Y}\sim 300$, which is typical for the
minijet gluon system produced at RHIC, we get the screen mass $\mu\sim 4.5
fm^{-1}$. So in the cascade calculation, the screen mass is taken as several
inverse fermis.

\section{General features of ZPC 1.0.1}

ZPC 1.0.1 provides the user with a choice of several scattering prescriptions,
the option to read in initial conditions or to generate some ideal initial
conditions, and the space cell division optimization.

\subsection{Scattering prescription}

Because of the geometric interpretation of the scattering cross section,
information travels across the distance of closest approach when a scattering
happens in a time period of the fixed time step or the mean free time. This may
lead to causality violation. Different prescriptions may give different
results \cite{kortmeyer1}. 

To specify the scattering scheme, we need to specify the collision frame, the
collision space-time point, the ordering frame, the ordering time. 
The collision frame is the frame in which the collision happens. The ordering
frame is the frame in which the collisions are ordered in time. It is the
global frame where the observer is in. The ordering time is the time of a
collision that we use for ordering the collisions. 

In ZPC 1.0.1, the collision frame and space time point choice is set through
the first digit of the variable iord\_sch which can be specified in input
file zpc.ini. The second digit of iord\_sch gives the ordering time. The
ordering frame can be chosen by setting the variable ibst\_flag.

\subsection{Geometric configurations and built-in initial conditions}

ZPC 1.0.1 provides 3 different geometric configurations to do cascade. We can
study 3 dimensional expansion, 1 dimensional expansion and particles in a box
by setting the variable i\_config. Setting igen\_flag gives some built-in
initial conditions.

For the 3-d expansion, the built-in initial condition is like the
minijet gluon system generated in RHIC Au on Au central collisions.
4000 particles are uniformly distributed in the pseudo-rapidity range $-5$ to
$+5$. They are generated in local thermal equilibrium with a temperature
$500\;MeV$ ($\hbar = c = 1$). At $t=0$ they are within a transverse disk of
radius $5\;fm$. The particles are formed with a longitudinal formation time
$\tau_0 = 0.1\;fm$.

The 1 dimensional expansion built-in initial condition is different from the
3-d one in that the transverse positions of particles at their formation times
are within a square with x (y) goes from $-5\times size1\;(2)$ to
$+5\times size1\;(2)$.

The particles in a box built in initial conditions give particles in thermal
equilibrium with temperature $500\;Mev$. The particles are confined within a
box with x (y,z) goes from $-5\times size1\;(2,3)$ to $+5\times size1\;(2,3)$.

\subsection{Optimization for fast cascade code}

The parton cascade is different from the hadronic cascade for relatively low 
energy nuclear physics in that there are much more particles to 
start with in one event. For example, at RHIC energies, each central 
Au on Au event has around 4000 minijet gluons. To get the next collision 
partner for a particle after the collision, a number of N (N is the total 
number of particles) operations have to be performed. The total number of 
operations for one event is proportional to $N^2$. For $N\approx 4000$, it's 
difficult to get enough statistics for detailed study of events, e.g., 
correlations \cite{gyulassy2}. To improve the speed of the 
parton cascade, we chose to divide the space into cells. In this manner, to 
update a particle's collision partner, only the particles in the same and 
neighboring cells need to be checked. This makes the total number of 
operations go down to C*N*n (n is the average number of particles in one 
cell). 

Cells expand with time after time size (which is the minimum of the cell
dimensions size1, size2, and size3). But to ensure the results are exactly
the same as the results without cell division, size should be larger than the
interaction length. For reference, the user may want to produce the results
without cell division for reference and cross check. For the 3-d expansion
case, set $v1\;=\;v2\;=\;v3\;=\;size1\;=\;size2\;=\;size3\;=\;0$; 
for 1-d expansion, set $i\_config\;=\;3$; for particles in a box set 
$i\_config\;=\;5$. 

\subsection{Optional OSCAR standard output}

OSCAR (Open Standard Codes and Subroutines) \cite{oscar1} standard has been
proposed by Yang pang and passed by a group of physicists (Open Standard
working Group) in the field of cascade simulation of high energy
nucleus-nucleus collisions to provide objective scientific criteria for cascade
simulations. Currently, the standard output (OSCAR1997A) has a file-header and
an event-header. The file-header contains general code and simulation
information and the event header stores particle number, particle ID, 4
momentum, mass, position and time. We can set $i\_oscar=1$ from file zpc.ini to
write out the standard output into the file zpc.oscar. This provides a way of
reconstructing event history.

\section{Program description}
\subsection{Main subroutines}

To clearly analysis the program structure, several technical words must be
defined in the ZPC context. An {\it EVENT} means the time evolution of a
particular initial phase space distribution of particles. For one particular
event, the user can test the effect of different random number sequences by
running the same initial conditions with different random number sequences. The
different space time evolutions with one same initial phase space distribution,
different random number sequences are called different {\it RUN}s. An {\it
  OPERATION} is one of the following: the formation of a particle, the
collision of a particle with cell boundary, the collision of two particles or a
combination of these simple operations. {\it PARTICLE INDEX} is the label of
particle in the common block /prec2/ and {\it PARTICLE NUMBER} is the
particle label in the common block /prec1/. The following
main subroutines are all in the main program zpc and control the flow of the
program.

\begin{itemize}

\item SUBROUTINE INI\_ZPC \\
Purpose: to initialize the program. It reads in the parameters
from the initialization file zpc.ini and initializes the main common
blocks.\\

\item SUBROUTINE INI\_EVT \\
Purpose: to initialize one particular event. It reads in
initial conditions for an event and, or generates initial conditions depending
on the user's requirement. Then it boosts the initial conditions if the user
requires.\\

\item SUBROUTINE INI\_RUN \\
Purpose: to reset the initial conditions to the same as last run.\\

\item SUBROUTINE ZPC\_RUN \\
Purpose: to select the next operation according to ordering time,
perform the operation and update the interaction list. \\

\item SUBROUTINE ZPC\_ANA1 \\
Purpose: to do analysis after each operation.  This
enables us to look at the time development of physical observables.\\

\item SUBROUTINE ZPC\_ANA2 \\
Purpose: to analyze event final states.\\

\item SUBROUTINE ZPC\_OUT \\
Purpose: to average results over events and write the results into
output files.\\

\end{itemize}

\subsection{Random number generator(s)}
Default random number generator in ZPC-1.0.1 is the linear congruential method
random number generator ran1 \cite{press1}. Another random number generator
ran3 based on the subtractive method is also included. 

\subsection{Main common blocks}

There are seven common blocks in ZPC-0.2.1 that contain parameters for the
cascade simulation. They are named para*, where * is a number going from 1 to
7.

\begin{itemize}

\item  COMMON /PARA1/ MUL \\
MUL: the multiplicity of one particular event.

\item  COMMON /PARA2/ XMP, XMU, ALPHA, RTS\_CUT2, CUTOFF2 \\ 
XMP: the mass of the particles. In ZPC 1.0.1, there is only one
type of particle, the gluon.\\
XMU: the screening mass.\\
ALPHA: the strong interaction coupling constant. \\
RTS\_CUT2 : lower square root of s cut for parton collisions.\\
CUTOFF2: the total parton scattering cross section divided by $\pi$.

\item  COMMON /PARA3/ N\_STA\_EVT, N\_EVENT, N\_RUN, I\_EVENT, I\_RUN  \\
N\_STA\_EVT: the starting event number in the input file. \\
N\_EVENT: the total number of events. \\
N\_RUN: the total number of runs per event. \\
I\_EVENT: the current event number that goes from 1 to N\_EVENT.\\
I\_RUN: the current run number.

\item  COMMON /PARA4/ IFT\_FLAG, IRE\_FLAG, IGEN\_FLAG, IBST\_FLAG  \\ 
IFT\_FLAG: switch for the generation of particle formation time.\\
= 0: generate formation time. \\
= 1: read in formation time from zpc.inp.\\ 
IRE\_FLAG: switch to read initial conditions from zpc.inp. \\ 
= 0: read the initial conditions.\\
= 1: do not read the initial conditions.\\ 
IGEN\_FLAG: switch to generate initial conditions.\\
= 0: do not generate initial conditions.\\ 
= 1: generate initial conditions.\\
IBST\_FLAG: choice of global frame.\\
= 0: the global frame (the ordering frame) is the collider center of mass
 frame for Au on Au collisions. \\ 
= 1: means the global frame is taken to be the target frame. 

\item  COMMON /PARA5/ I\_CONFIG, IORD\_SCH \\ 
I\_CONFIG: choice of geometric configuration and space cell division
optimization.\\ 
= 1: the system is undergoing 3-d expansion.\\
= 2: the system is undergoing 1-d expansion with space cell division.\\
= 3: the system is undergoing 1-d expansion without space cell division.\\
= 4: the system is confined in a box with space cell division.\\
= 5: the system is confined in a box without space cell division.\\
IORD\_SCH: choice of collision scheme. IORD\_SCH = 10 * I1 + I2\\
I1: choice of collision prescription.\\
= 0: the collision frame is the two parton center of mass frame. The scattering
space point is the center point of the 2 parton positions at closest approach
in the collision frame.\\
= 1: the collision frame is the two parton center of mass frame. The scattering
point is the position of the parton at the closest approach in the collision
frame. This gives two generally different collision
times for the two colliding partons.\\
= 2: the collision frame is the same as the ordering global frame.\\
I2: choice of ordering time.\\
= 0: the earlier of the two collision times in the global frame is the ordering
time for parton collisions.\\
= 1: the average of the two collision times is taken as the ordering time.\\
= 2: the later collision time is the ordering time.

\item  COMMON /PARA6/ CENT\_RAP \\
CENT\_RAP: the central rapidity of the rapidity plateau.

\item  COMMON /PARA7/ I\_OSCAR \\
I\_OSCAR: option to write OSCAR standard output.\\
= 0: write OSCAR standard output.\\
= 1: do not write OSCAR standard output.

\end{itemize}

Six particle record common blocks are labeled prec!, in which ! goes from 1 to
6. 

\begin{itemize}

\item  COMMON /PREC1/ \\
Purpose: stores the initial conditions for one event.

\item  COMMON /PREC2/ \\
Purpose: main particle record that has particle types, the positions at
         formation times, 
         the corresponding formation times and the 
         four momenta of the particles, plus the masses of the particles.

\item  COMMON /PREC3/ \\
Purpose: the same as /PREC2/ except the two colliding 
        particles don't update their records until next collision.

\item  COMMON /PREC4/ \\
Purpose: stores velocities of particles.

\item  COMMON /PREC5/ \\
Purpose: stores formation time pseudo-rapidities, rapidities and proper times
         of particles. 

\item  COMMON /PREC6/ \\
Purpose: the same as /PREC5/ except the two colliding 
        particles don't update their records until next collision.

\end{itemize}

Six interaction list common blocks i\_list! give us particle run time
information and cell information. 

\begin{itemize}

\item  COMMON /I\_LIST1/ ISCAT, JSCAT, NEXT(NMAXGL), LAST(NMAXGL),
ICTYPE, ICSTA(NMAXGL), NIC(NMAXGL), ICELSTA(NMAXGL) \\
ISCAT: particle index. If the operation involves only one particle,
 then iscat is the particle index; if it involves 2
 particles, iscat is the larger particle index.\\
JSCAT: particle index. If the operation involves only one particle,
 then jscat is 0; if it involves 2 particles, jscat is the smaller
 particle index.\\
NEXT(I): the next operation partner of particle i.\\
LAST(I): the last operation partner of particle i.\\
ICTYPE: the operation type.\\ 
= 0: a collision between particles.\\ 
= 1: the formation of a particle.\\ 
= 2: both a collision between particles and the formation of a particle.\\ 
= 3: a collision with wall.\\ 
= 4: both a collision between particles and a wall collision.\\ 
= 5: both a wall collision and a formation.\\ 
= 6: a formation, collision between particles and a wall collision at
 the same time.\\ 
ICSTA(I): the operation type for particle i.\\ 
= 0: an ordinary collision.\\ 
= 101: a collision with the wall with larger x.\\
= 102: a collision with the wall with smaller x.\\
= 103: a collision with the wall with larger y.\\ 
= 104: a collision with the wall with smaller y.\\
= 105: a collision with the wall with larger z.\\ 
= 106: a collision with the wall with smaller z.\\
= 111: a collision with another particle and the wall with larger x.\\ 
= 112: a collision with another particle and the wall with smaller x.\\ 
= 113: a collision with another particle and the wall with larger y.\\ 
= 114: a collision with another particle and the wall with smaller y.\\ 
= 115: a collision with another particle and the wall with larger z.\\ 
= 116: a collision with another particle and the wall with smaller z.\\ 
NIC(I): the next particle index in the same cell as i.\\
ICELSTA(I): the encoded information of the cell number particle i is in.
 If particle is in cell (i1, i2, i3), then it equals $i1\times 10000 +
 i2\times 100 + i3$ for particle inside the cube. When a particle is outside
 the 10 by 10 by 10 box, its value is $111111$.

\item  COMMON /I\_LIST2/ ICELL, ICEL(10,10,10) \\
ICELL: pointer to one particle out side the 10 by 10 by 10 cell volume. \\
ICEL(I, J, K): one particle in the cell (i, j, k).

\item  COMMON /I\_LIST3/ SIZE1, SIZE2, SIZE3, V1, V2, V3, SIZE \\
SIZE1: cell size in the x direction. It must be larger than the inverse square
root of cutoff2.\\ 
SIZE2: cell size in the y direction.\\
SIZE3: cell size in the z direction.\\
V1: cell expanding velocity in x direction.\\
V2: cell expanding velocity in y direction.\\
V3: cell expanding velocity in z direction.\\
SIZE: the time that cells begin to expand.

\item  COMMON /I\_LIST4/ IFMPT, ICHKPT, INDX(NMAXGL) \\
IFMPT: the particle index of the parton that is to be formed next.\\
ICHKPT: the last formed particle index.\\
INDX(I): the index of particle i of /INI\_REC/ in an increasing formation time
order.

\item  COMMON /I\_LIST5/ CT(NMAXGL), OT(NMAXGL), TLARGE \\
CT(I): the collision time for particle i.\\
OT(I): the ordering time for particle i.\\
TLARGE: a large number for time cutoff.

\item  COMMON /I\_LIST6/ I\_OPERATION, I\_COLLISION, T \\
N\_OPERATION: the current number of operations.\\
I\_COLLISION: the current number of collision between particles that have been
        performed.\\
T: the current operation time.

\end{itemize}

\subsection{Other common blocks}

Two common blocks /AUX\_REC1/ and /AUX\_REC2/ contain auxiliary variables for
1-d expansion and particles in a box without space cell division.

Two common blocks contain information about the random numbers.

\begin{itemize}

\item  COMMON /RANDOM1/ NUMBER  \\
NUMBER: the number of random numbers that have been used.

\item  COMMON /RANDOM2/ IFF \\
IFF: choice of attractive or repulsive force. it alternates between 1 and -1.

\end{itemize}

Currently, there are four common blocks ana! for analysis.

\begin{itemize}

\item  COMMON /ANA1/ TS(12) \\
TS: stores the time points that we want to sample data for analysis.

\item  COMMON /ANA2/ \\
Purpose: to record $dE_T/dy$, $dE_T/dN$, and $dN/dy$ time evolution.

\item  COMMON /ANA3/ EM(4, 4, 12) \\
EM: energy momentum tensor for 12 different time points.

\item  COMMON /ANA4/ FDETDY(24), FDNDY(24), FDNDPT(12) \\
FDETDY: final $dE_T/dy$ distribution.\\
FDNDY: final $dN/dy$ distribution. \\
FDNDPT: final $dN/dp_T$ distribution.

\end{itemize}

\subsection{Input files}

\begin{itemize}

\item {\bf zpc.ini} \\
Purpose: provides the following parameters for ZPC:
\subitem 'mass of particles (GeV)'
\subitem 'screening mass (fm\^(-1))'
\subitem 'alpha'\\
       strong interaction coupling constant;
\subitem 'rts\_cut2 (GeV\^2)'\\
     square root s cutoff bellow which no collision will by taken into account;
\subitem 'n\_sta\_evt'\\
      starting event number;
\subitem 'n\_event'\\
      number of events;
\subitem 'n\_run'\\
      number of runs;
\subitem 'gene ft flag (0,yes;other,no)';
\subitem 'read ini flag (0,read;other,no)';
\subitem 'gene ini flag (0,no;other,yes)';
\subitem 'boost ini condition (0,no)';
\subitem 'i\_config (1,3-d exp;2,3 1-d exp;4,5 box)'
\subitem 'collision ordering scheme' gives the parameter IORD\_SCH;
\subitem 'optional OSCAR output: (0,no;1,yes)'
\subitem 'v1, v2, v3'\\
  v1, v2, v3 are the cell expanding velocities;
\subitem 'size1, size2, size3 (fm)' \\
 size1, size2, and size3 give the cell sizes in the x, y, z directions;
\subitem 'beginning iff'\\
  beginning iff gives the beginning random choice of attractive or repulsive
  collision;
\subitem 'iseed'\\
  iseed is the random number seed;
\subitem 'i\_ran\_used' \\
  number of random numbers used. This makes it possible to continue a series of events.

\item {\bf zpc.inp} \\
Purpose: provides initial conditions for ZPC. Each particle has one
record in zpc.inp. Each record has 11 entries: the event number,
the particle type (integer), the x, y, and
z positions of the particle at the time specified, the specified time, and the
four momentum of the particle, the mass of particle (double precision). 
The specified time can be one time same for
all the particles, in this case, the ift\_flag in the zpc.ini file should be
set to 0. The specified time can also be the given formation time of the
particle. In this case, the ift\_flag should be set to 1.

\end{itemize}

\subsection{Output files}
\begin{itemize}

\item {\bf zpc.res} \\
zpc.res is the major output file. It has the event number, run
number, number of operations, number of collisions between particles,
freezeout time, ending random number, and ending randomizer iff.The user can
dump other analysis information into zpc.res.

\item {\bf zpc.jun} \\
zpc.jun saves execution time and error message.

\item {\bf zpc.oscar} \\
ZPC 1.0.1 writes OSCAR standard output \cite{oscar1} into zpc.oscar.

\item {\bf ana1 directory}  \\
Currently files in ana1 are for the Lorentz invariance analysis.
fdetdy.dat is the final detdy distribution; fdndy.dat is the final dndy
distribution; fdndpt.dat is the final dndpt distribution.

\item {\bf ana2 directory} \\
Currently files in ana2 are for the time evolution of detdy dndy
and detdydn analysis. detdy*.dat, dndy*.dat, and detdydn*.dat are for the time
evolution of detdy, dndy, detdydn correspondingly. * is a number that specifies
the rapidity range of the central slice that the program takes data from.

\item {\bf ana3 directory} \\
ana3 has information for 1-d expansion semi-analytic tests. coll.res has the
collision proper time and Mandelstam s for central collisions. 
et*.res files give lab time development of $dE_T/dy$ at fixed proper
times. et*.res files are useful in choosing a final freezeout time for 1-d
expansion such that the $dE_T/dy|_{y=0}(\tau)$ reaches asymptotic value. 

\item {\bf ana4 directory} \\
ana4 contains energy-momentum tensor at different times for particles in a box
case. 

\end{itemize}

\subsection{Other files}
\begin{itemize}

\item {\bf README} \\
README gives a brief description of the files of ZPC.

\item {\bf COPYING} \\
COPYING contains GNU General Public License.

\item {\bf zpc.doc} \\
zpc.doc has detailed discussion of the program and instruction on
running the program.

\item {\bf zpc.f}  \\
zpc.f is the main source file.

\item {\bf zpc.go}  \\
zpc.go is the script file that controls the running of the program.
On some machines, there are special rules for batch jobs, zpc.go must be
modified accordingly.

\item {\bf zpc.mk}  \\
zpc.mk is the makefile that generates the executable file.

\end{itemize}

\section{Instructions for users}

\begin{itemize}

\item Set parameters in zpc.ini;
\item Make zpc.inp if the initial conditions are to be read in from it; 
\item Run the program by running zpc.go. It will use zpc.mk to make the 
    executable file, run and time the executable file.  

\end{itemize}

Here is an example input file and corresponding results.

Input file zpc.ini:

\begin{verbatim}
'mass of particles (GeV)                  '      0d0
'screening mass (fm^(-1))                 '      3d0
'alpha                                    '      0.47140452
'rts_cut2 (GeV^2)                         '      0.01d0
'n_sta_evt                                '      1
'n_event                                  '      10 
'n_run                                    '      1
'gene ft flag (0,yes;other,no)            '      1
'read ini flag (0,read;other,no)          '      1
'gene ini flag (0,no;other,yes)           '      1
'boost ini flag (0,no;other,yes)          '      0
'i_config (1,3-d exp;2,3 1-d exp;4,5 box) '      1
'collision ordering scheme                '      11
'optional OSCAR output: (0,no;1,yes)      '      0
'v1, v2, v3                               '      0.2d0 0.2d0 0.2d0
'size1, size2, size3 (fm)                 '      1.5 1.5 0.7
'beginning iff                            '      -1
'iseed                                    '      7
'i_ran_used                               '      1

\end{verbatim}

Output file zpc.res:

\begin{verbatim}
Event  1, run  1,
         number of operations =   36224,
         number of collision between particles =   6914,
         freezeout time=    70119.524722901,
         ending at the   45159th random number,
         ending collision iff=  -1
Event  2, run  1,
         number of operations =   35886,
         number of collision between particles =   6856,
         freezeout time=    447736.85516922,
         ending at the   90145th random number,
         ending collision iff=  -1
Event  3, run  1,
         number of operations =   36118,
         number of collision between particles =   6954,
         freezeout time=    99267.891489987,
         ending at the   135343th random number,
         ending collision iff=  -1
...

\end{verbatim}

The data files in directory ana1 give the final $\frac{dE_T}{dy}$,
$\frac{dN}{dy}$ and $\frac{dN}{dp_T}$. They are shown in Fig. 1, Fig.2 and
Fig. 3.
\begin{figure}
\vspace{0.5cm}
\hspace{1.21cm}
\psfig{figure=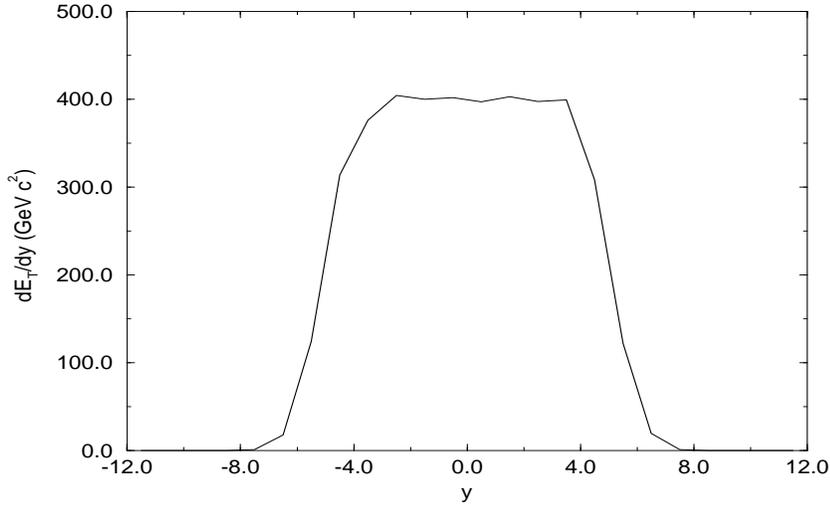,height=2.9in,width=2.6in,angle=-90}
\caption{10 events, $\mu=3\;fm^{-1}$}
\label{Fig. 1}
\end{figure}

\begin{figure}
\vspace{0.5cm}
\hspace{1.21cm}
\psfig{figure=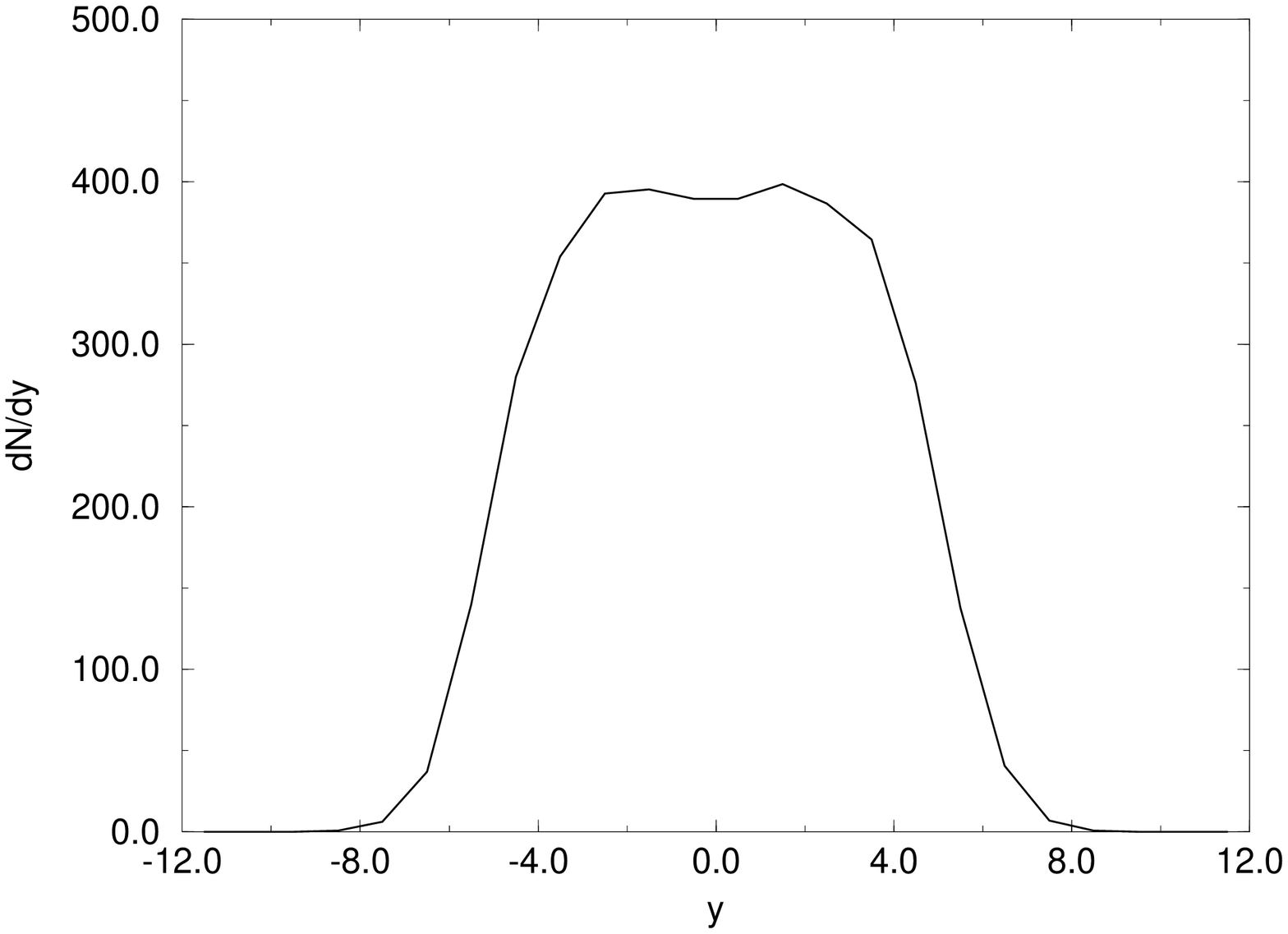,height=2.9in,width=2.6in,angle=-90}
\caption{10 events, $\mu=3\;fm^{-1}$}
\label{Fig. 2}
\end{figure}

\begin{figure}
\vspace{0.5cm}
\hspace{1.21cm}
\psfig{figure=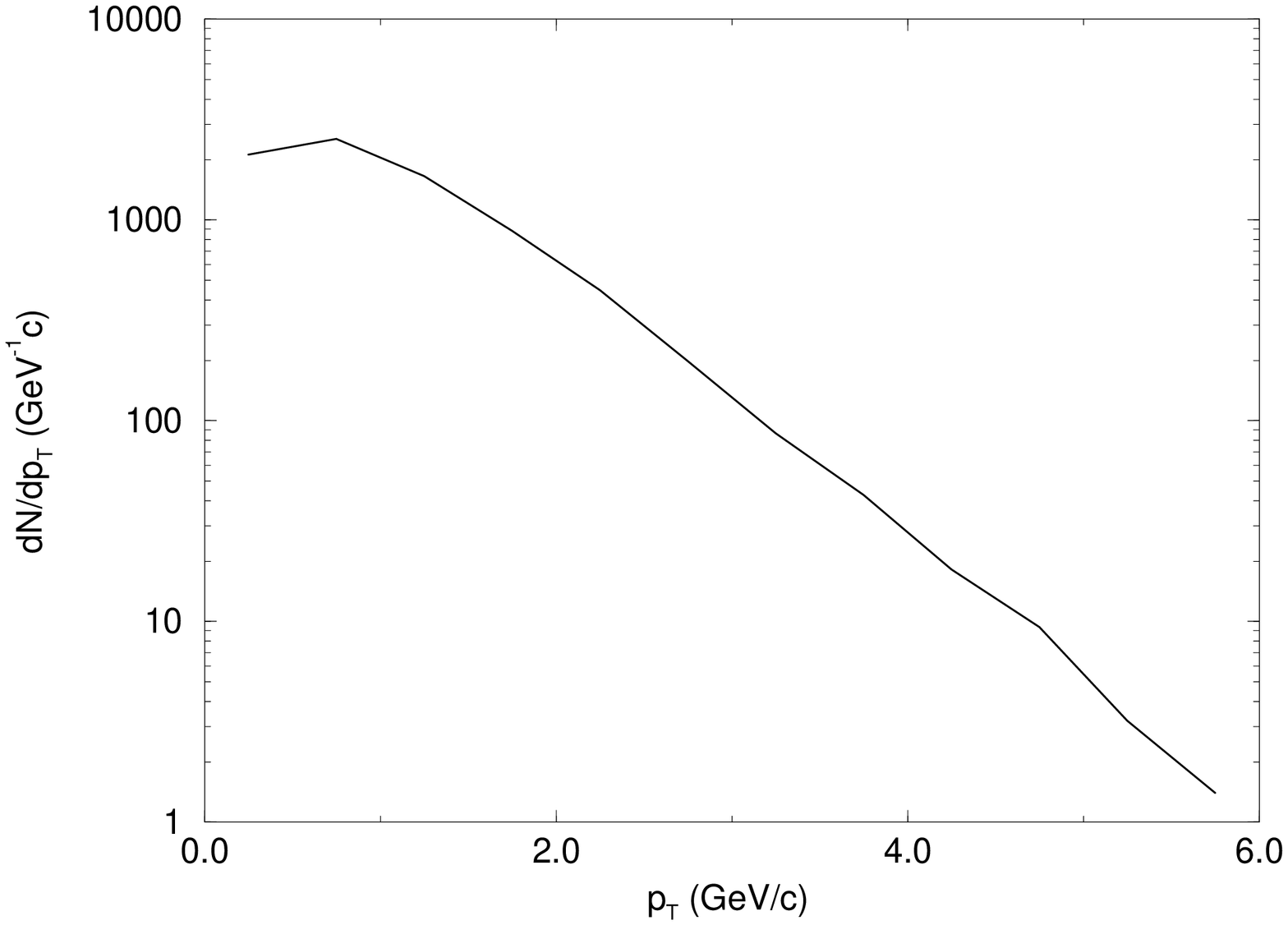,height=2.9in,width=2.6in,angle=-90}
\caption{10 events, $\mu=3\;fm^{-1}$}
\label{Fig. 3}
\end{figure}

\section{Modification and further development}
The current version 1.0.1 has only elastic collisions. It can be easily 
modified to study more complicated physical process, for example, to study
effects of elastic collisions on chemical equilibrium. To put in more particle 
species:

\begin{itemize}

\item Add in new parameters in zpc.ini and change subroutine ini\_zpc 
    correspondingly. 
\item Add in zpc.inp more entries corresponding to new degrees of freedom for 
    particles and change read\_ini correspondingly. 
\item In subroutine scat, new subroutine is necessary to update the new degrees     of freedom for the colliding particles. subroutine get\_that needs 
    to be changed correspondingly to get the scattering angle through the new 
    scattering differential cross section.
\item In subroutine iscoll*, new subroutine get\_cutoff2 is necessary to get 
    $\frac{\sigma}{\pi}$ for different processes.

\end{itemize}

Further analysis can be done by putting the analysis in the subroutine 
zpc\_ana1 and zpc\_ana2, and adding corresponding output in zpc\_out.\\[3ex]

{\Large \bf Acknowledgments}\\[1ex]

I'm very grateful to M. Asakawa, S. Bass, P. Danielewicz, S. Gavin, K. Geiger,
D. Kahana, S. Kahana, B. Li, Z. Lin,
R. Mattiello, C. Noack, D. Rischke, S. Pratt, J. Randrup, A. Tai,
K. Werner for  useful discussions. Special thanks go to X. N. Wang, Y. Pang and
M. Gyulassy for continuing encouragement and discussions and careful reading of
the manuscript. I'd also like to thank Brookhaven National Laboratory and
Lawrence Berkeley Laboratory for providing computing facilities.

{}

\end{document}